\documentclass[conference]{IEEEtran}
\pdfoutput=1

\usepackage{multirow}

\usepackage{cite}

\ifCLASSINFOpdf
   \usepackage[pdftex]{graphicx}
\else
\fi
\usepackage{amsmath}
\ifCLASSOPTIONcompsoc
  \usepackage[caption=false,font=normalsize,labelfont=sf,textfont=sf]{subfig}
\else
  \usepackage[caption=false,font=footnotesize]{subfig}
  \fi

\begin{document}

%
\title{Voltage Stability Studies for Distribution Networks: Assessing Load Dynamics}

\author{\IEEEauthorblockN{Ruth Kravis, Ian Petersen, Elizabeth Ratnam}
\IEEEauthorblockA{School of Engineering, The Australian National University, Canberra, Australia\\
Email: ruth.kravis@anu.edu.au; ian.petersen@anu.edu.au; elizabeth.ratnam@anu.edu.au}
}


%


\maketitle

\begin{abstract}
This paper presents an investigation into load dynamics that potentially cause voltage instability or collapse in distribution networks. Through phasor-based, time domain simulations of a dynamic load (DL) model from the literature, we show that the load dynamics alone do not cause voltage instability or collapse. By comparing the DL model to a benchmark model, we identify an important limitation with the DL model. We characterise this limitation and recommend that future work use load models with physical state variables. By investigating when and how load dynamics cause voltage instability, we are well-positioned to develop systems to control and maintain voltage stability in distribution networks.

\end{abstract}


%

\section{Introduction}


 



The recent rapid increase in distributed generation and storage within distribution networks has created concern for maintaining voltage stability across both transmission and distribution networks. Voltage stability refers to the ability of a power system to recover from a disturbance and return to a steady-state voltage operating condition that is within upper and lower bound operating thresholds \cite{IEEE_CIGRE_Definitions}. Ensuring voltage stability is crucial for grid resilience. If voltage instability is not detected and corrected, it can lead to voltage collapse, where the system voltage uncontrollably decays. In addressing this challenge, we must use appropriate models that capture the important dynamic behaviour of devices including loads and distributed energy resources (DER). \\

Historically, voltage instability and collapse events have occurred at the transmission level, driven by the dynamics of large generators and the maximum power transfer limits of the power system \cite{Kundar_textbook}. The behaviour of loads have been considered secondary, with static load models commonly used in voltage stability studies. Studies that have focused on the impact load modelling choices have on voltage stability assessments have shown that different load models give different results \cite{1994_Overbye,Borghetti1997}. Often, these studies only consider steady-state voltage stability, relying on linearised system equations. In these studies, the stability limit is given by the point at which the power-flow Jacobian is singular \cite{1994_Overbye}. \\

However, field tests on real distribution networks have confirmed that individual loads, and aggregations of many small load devices, have dynamic, non-linear behaviour \cite{Xu,Hill_LoadModel,MKPal_stability_load}. To study large disturbance voltage stability, or even voltage stability in heavily loaded systems, it is important to use models that capture the load's non-linear, dynamic behaviour. 

As DER proliferation continues, electrical networks are expected to operate closer to their physical limits reducing the need for grid expansion. Consequently, the importance of load dynamics, and how the dynamics influence voltage stability will become increasingly important \cite{Hiskens_tools_nonlinearities}. However, much of the existing literature is focused on voltage stability in transmission networks which does not apply directly to distribution grids as the assumption that line impedances have negligible resistance $R\neq0$ does not hold \cite{PBC_energies}. Furthermore, existing studies have not clearly indicated the possibility of load dynamics driving a system to instability under distribution network assumptions \cite{Hill_LoadModel,Xu,MKPal_stability_load,Hill_dynamic_analysis_voltage_collapse}. Thus, an investigation into the limitations of existing dynamic load models in the context of distribution networks addresses a significant gap in the literature. \\

In this paper, we examine a dynamic load (DL) model reported in \cite{Xu}, and compare its performance to a benchmark model. Simulations of non-steady state behaviour and disturbance response are conducted, for a range of plausible system parameters. The analysis conducted shows that the dynamic load model can only provide a limited explanation of what voltage instability looks like in distribution networks. The model cannot be used to observe the system's dynamic behaviour outside the stable region of the power-flow equations, even when the system has an analytical low voltage equilibrium that is stable under different modelling choices. \\

This paper is organised as follows. In Section \ref{sec:problem_formulation}, we define the DL model and the test system, as well as the benchmark model. In Section \ref{sec:results} we present our simulation results and discuss our observations in the context of voltage stability in distribution networks. Finally, conclusions and recommendations are presented in Section \ref{sec:conclusion}.

\section{Problem Formulation}\label{sec:problem_formulation}
\subsection{Dynamic Load (DL) Model for Distribution Networks}

\begin{figure}[!h]
   \centering
   \includegraphics[width=.8\linewidth]{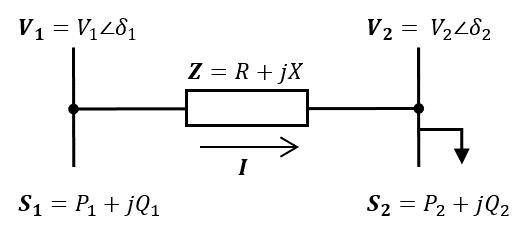}   
    \caption{Two-node distribution circuit, with Node 1 sending power and Node 2 receiving power.}\label{fig:DL_circuit} 
\end{figure}

We model a two-node distribution circuit, with an infinite bus at Node 1, and a load connected at Node 2. The two-node distribution circuit is presented in Fig~\ref{fig:DL_circuit}. 

We use phasor representation throughout. We denote by $V_1$ and $\delta_1$ the voltage magnitude and voltage phase angle at Node 1, and by $V_2$ and $\delta_2$ the voltage magnitude and voltage phase angle at Node 2. We denote by $R$ and $X$ the line resistance and reactance respectively, and by $P_2$ and $Q_2$ the real and reactive power respectively at Node 2. Assuming $\delta_1=0$ and $R=X$, the algebraic constraints defining power flow between Node 1 and 2 are given by:

\begin{align}
    \begin{split}
    V_2^4 +((2RP_2+2RQ_2)-V_1^2)V_2^2 + \\
    2R^2P_2^2+2R^2Q_2^2=0,\label{eqn:eq1}
    \end{split}
\end{align}

\begin{equation}
\delta_2= -\frac{\pi}{4} + cos^{-1}\left(\frac{\sqrt{2}RP_2+(\sqrt{2}/2)V_2}{ V_1 V_2}\right).\label{eqn:2node_angles}
\end{equation}

We use the generic dynamic load model presented in \cite{Xu}. The load is modelled as a linear system $\dot{x}=Ax+B$, with the entries of $A$ and $B$ varying non-linearly with the voltage $V_2$. Equation~\eqref{eqn:load_dynamics} defines the state-space load model, and \eqref{eqn:Ps}--\eqref{eqn:Qt} define the steady state and transient load characteristics that are used as entries in the $A$ and $B$ matrices. 

\begin{align}
\begin{split}
\begin{bmatrix} 
\dot{x} \\
\dot{y}
\end{bmatrix}=\begin{bmatrix} 
-\frac{P_t(V_2)}{T_p} & 0 \\
0 & -\frac{Q_t(V_2)}{T_q} 
\end{bmatrix}\quad
\begin{bmatrix} 
x  \\
y
\end{bmatrix}\quad + 
\begin{bmatrix} 
1 & 0 \\
0 & 1
\end{bmatrix}\quad
\begin{bmatrix} 
\frac{P_{s}(V_2)}{T_p}  \\
\frac{Q_{s}(V_2)}{T_q} 
\end{bmatrix}
\end{split}
\label{eqn:load_dynamics}
\end{align}

\begin{align}
P_s(V_2):=P_{0}V_2^{a}, \label{eqn:Ps}\\ 
Q_s(V_2):=Q_{0}V_2^{b}, \label{eqn:Qs} \\
P_t(V_2) \approx -0.08V_2^2 +0.96V_2 +0.12, \label{eqn:Pt}\\ 
Q_t(V_2) \approx 3.255V_2^2-3.49V_2+1.155. \label{eqn:Qt} 
\end{align}

The system is at equilibrium when: 

\begin{align}
    P_2 = P_s(V_2) = x{P_t(V_2)},   \label{eqn:P2_eqm}\\
    Q_2 = Q_s(V_2) = y{Q_t(V_2)}.  \label{eqn:Q2_eqm}
\end{align}

and when \eqref{eqn:eq1} is satisfied for $P_2$ and $Q_2$ given by \eqref{eqn:P2_eqm} and \eqref{eqn:Q2_eqm}:

\begin{align}
\begin{split}
V_2^4 +(2P_{0}V_2^{a}(R + X{\frac{Q_0}{P_0}V_2^{b-a}})-V_1^2)V_2^2 + \\ ({P_{0}V_2^{a}})^2\left(1+\left(\frac{Q_0}{P_0}V_2^{b-a}\right)^2\right)(R^2 + X^2)=0. \label{eqn:eq1_load_eqm} 
\end{split} 
\end{align}

Observe that the stability of a given equilibrium point will depend on both the algebraic constraint from \eqref{eqn:eq1} and the load dynamics from \eqref{eqn:load_dynamics}. Due to the system's non-linearity, we choose to investigate stability via simulation rather than analytically. 

\subsection{Benchmark Model}

\begin{figure}[!h]
\centering
    \includegraphics[width=.6\linewidth]{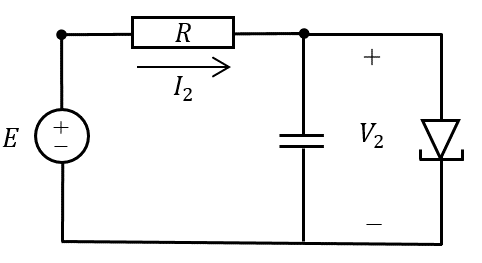} 
    \caption{Benchmark system circuit representation}
    \label{fig:TD_circuit}
\end{figure}

Equation~\eqref{eqn:eq1} in the DL model can have up to two positive, real voltage solutions. This is a potential source of ambiguity, so we formulate a DC benchmark model which always has a unique voltage solution. In Fig.~\ref{fig:TD_circuit} we present the circuit representation of this benchmark model, re-constructed from \cite{khalil2002nonlinear}. 

This circuit has a source voltage $V_1$, similar to the infinite bus in the previous model, and a line resistance $R$. The `load' is comprised of a capacitor in parallel with a tunnel diode. The capacitor has linear dynamic behaviour, and the tunnel diode has static non-linear behaviour. Together, they give nonlinear, dynamic behaviour. The system's behaviour is given by: 

 \begin{align}
    \dot{V_{2}} =\left[-h(V_2) - \frac{V_2}{R}+ \frac{V_1}{R}\right]\frac{1}{C}.
    \label{eqn:TD_differential_eqn_DC}
\end{align}

where $h(V_2) =aV_2^5 + bV_2^4 + cV_2^3 + dV_2^2 + eV_2$. The parameters of $h(V_2)$ are chosen such that the system has three equilibrium points, where an equilibrium point is a voltage satisfying the equation:

\begin{align}
    \frac{V_1 - V_2}{R} = h(V_2).
\end{align}

\begin{figure}[!h]
    \centering
    \includegraphics[width=.65\linewidth]{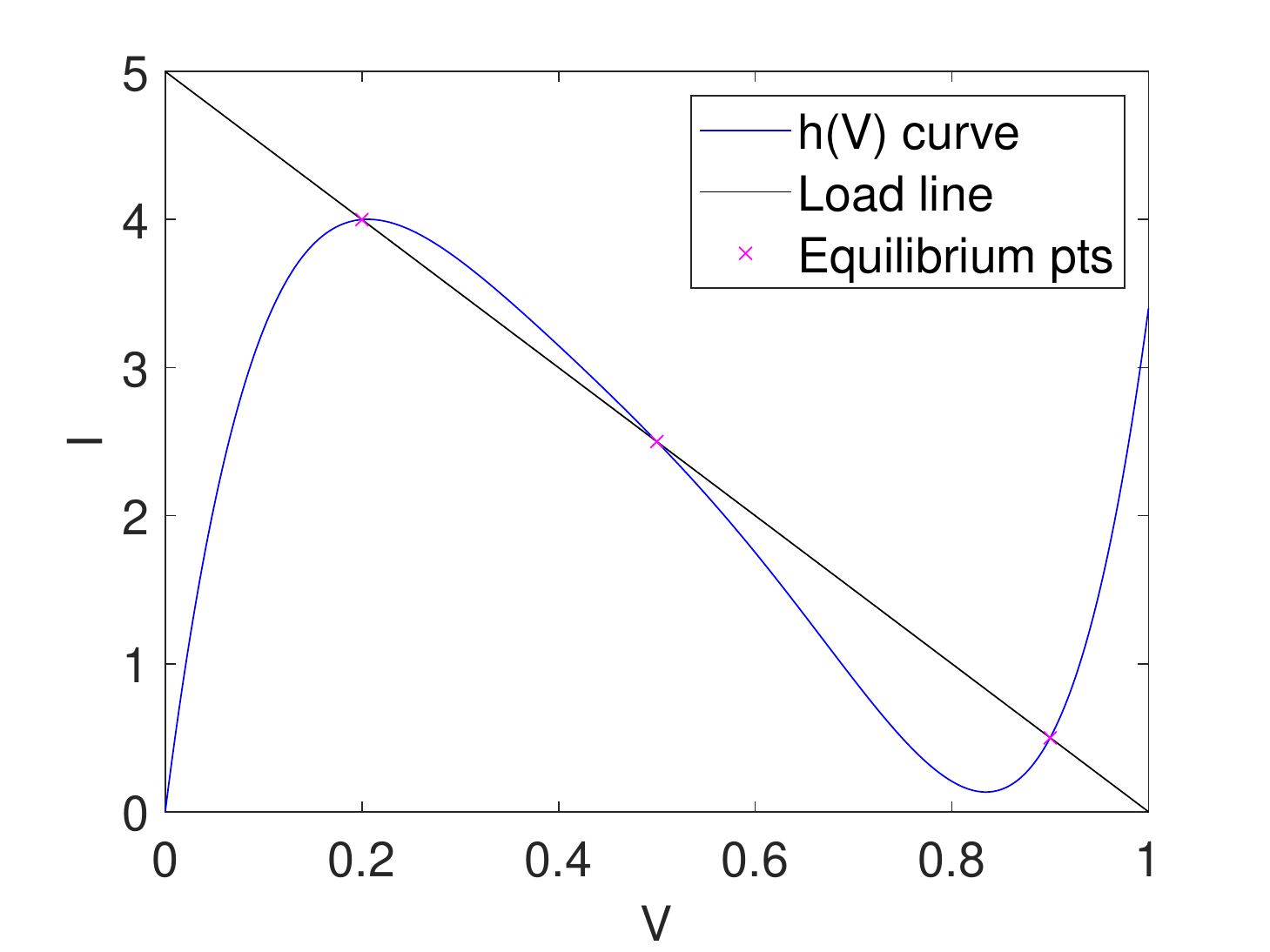}
    \caption{Benchmark model voltage-current characteristic}
    \label{fig:TD_VIP_xstics}
\end{figure}

Example circuit characteristics are provided in Fig.~\ref{fig:TD_VIP_xstics}, with the three equilibrium points marked. Due to the slope of $h(V_2)$ at the points of intersection with the Node 2 load line, the lower and upper equilibrium points are stable, and the middle equilibrium point is unstable. The stable equilibrium points each have well-defined regions of attraction, separated by the unstable middle equilibrium voltage. Furthermore, under a disturbance that causes a high voltage stable equilibrium point to be lost, the dynamics allow the system to transition from a high voltage to a low voltage equilibrium point. Conceptually, this benchmark model behaves as we would expect a physical power system to behave.

\section{Results}\label{sec:results}

The results presented in this section were obtained by running time-domain simulations in Simulink. A Runge-Kutta fixed step ODE solver with a timestep of 1 ms was used. All units are in p.u. unless otherwise stated.

\subsection{Benchmark: Dynamic simulations}
\begin{figure}[!h]
\centering
    \includegraphics[width=.8\linewidth]{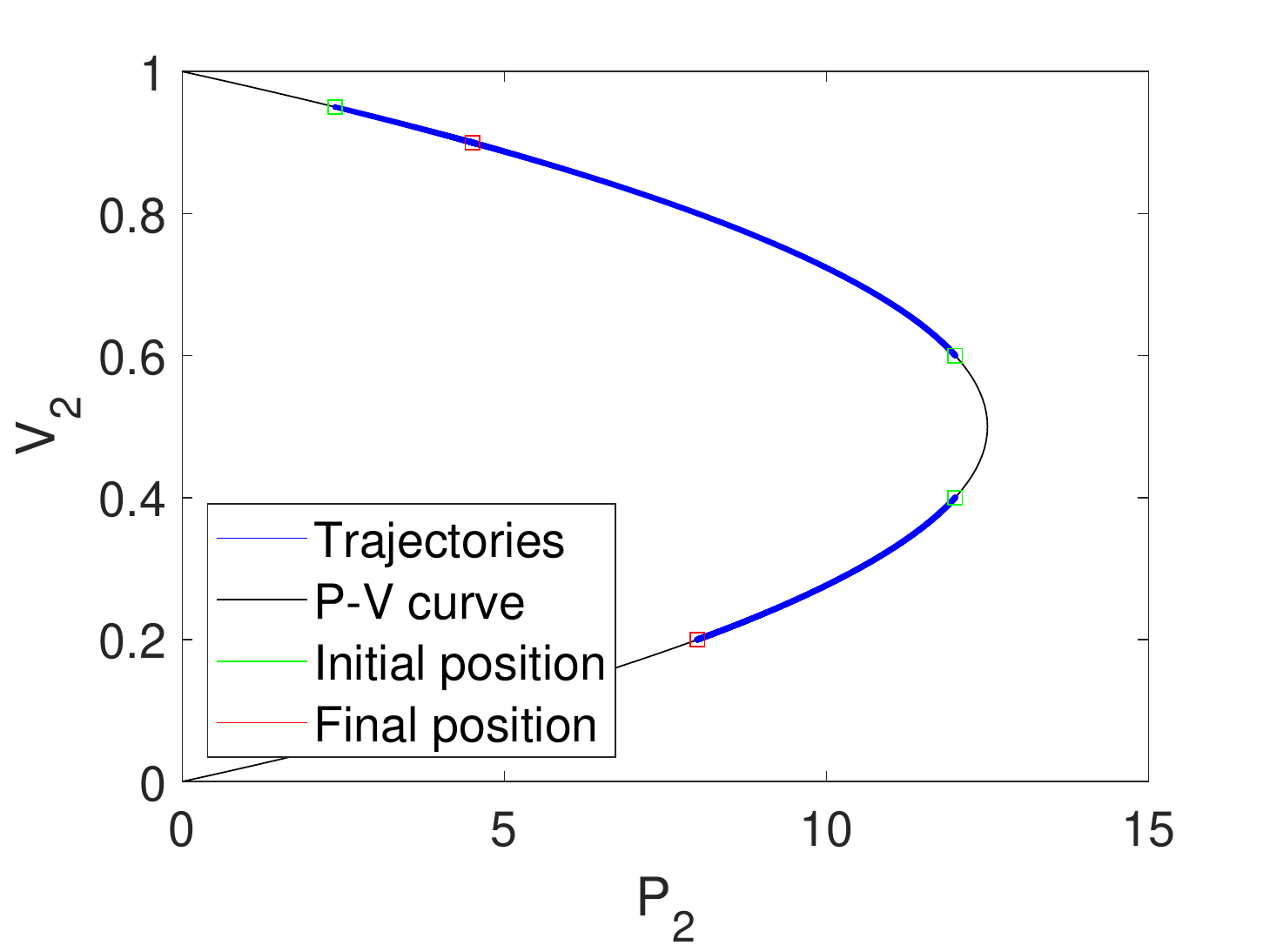} 
    \caption{TD trajectories for the system with parameters in Table \ref{tab:TD_parameters}. This system has two stable equilibrium points at $V_2=0.2, 0.9$, and one unstable equilibrium at $V_2=0.5$.}
    \label{fig:TD_trajs}
\end{figure}

Here, we present our investigation into the system response to a range of initial voltages for $V_2$. In Fig.~\ref{fig:TD_trajs} we present our simulation results for the benchmark system with parameters specified in Table \ref{tab:TD_parameters}. 

\begin{table}[!h]
    \centering
    \caption{Benchmark model parameters}
    \label{tab:TD_parameters}
    \begin{tabular}{cc}
    \hline
        Parameter & Value \\ \hline
        $V_1$ & 1 \\ 
        R    & 0.02 \\ 
        C & 10 \\ 
        a & 218.6576  \\ 
        b & -539.0593  \\ 
        c & 517.9071  \\
        d & -246.6894  \\ 
        e &  52.5842 \\ \hline
 \end{tabular}
\end{table}

From the three trajectories presented in Fig.~\ref{fig:TD_trajs} we can observe the regions of attraction for each stable equilibrium point. We observe that under certain initial $V_2$, the system converges to one stable equilibrium rather than the other. \\

\begin{figure}[!h]
\centering
\subfloat[P-V plane]{\includegraphics[width=.7\linewidth]{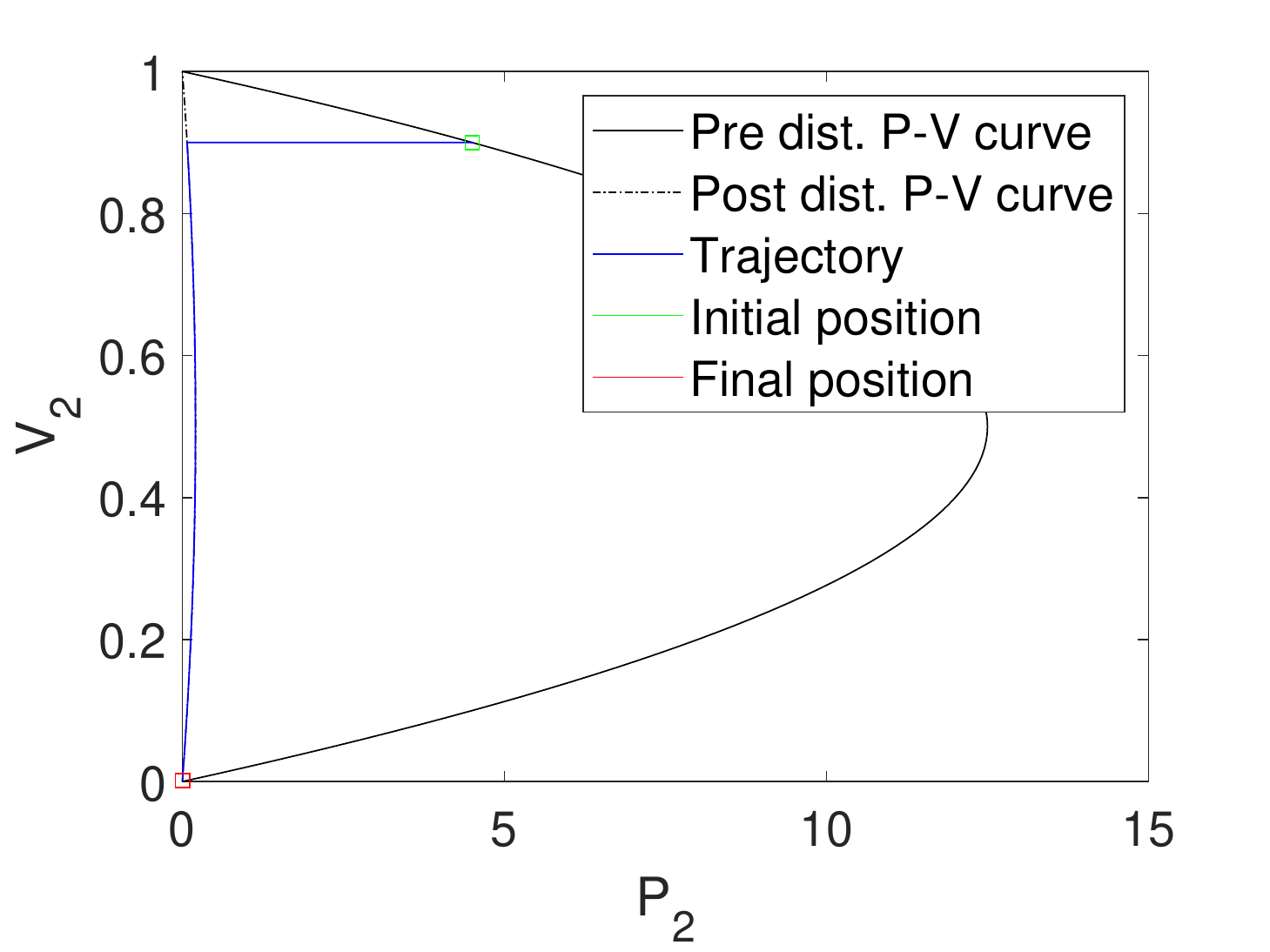}%
\label{TD_dist_PV}}
\hfill
\subfloat[$V_2$ time series]{\includegraphics[width=.7\linewidth]{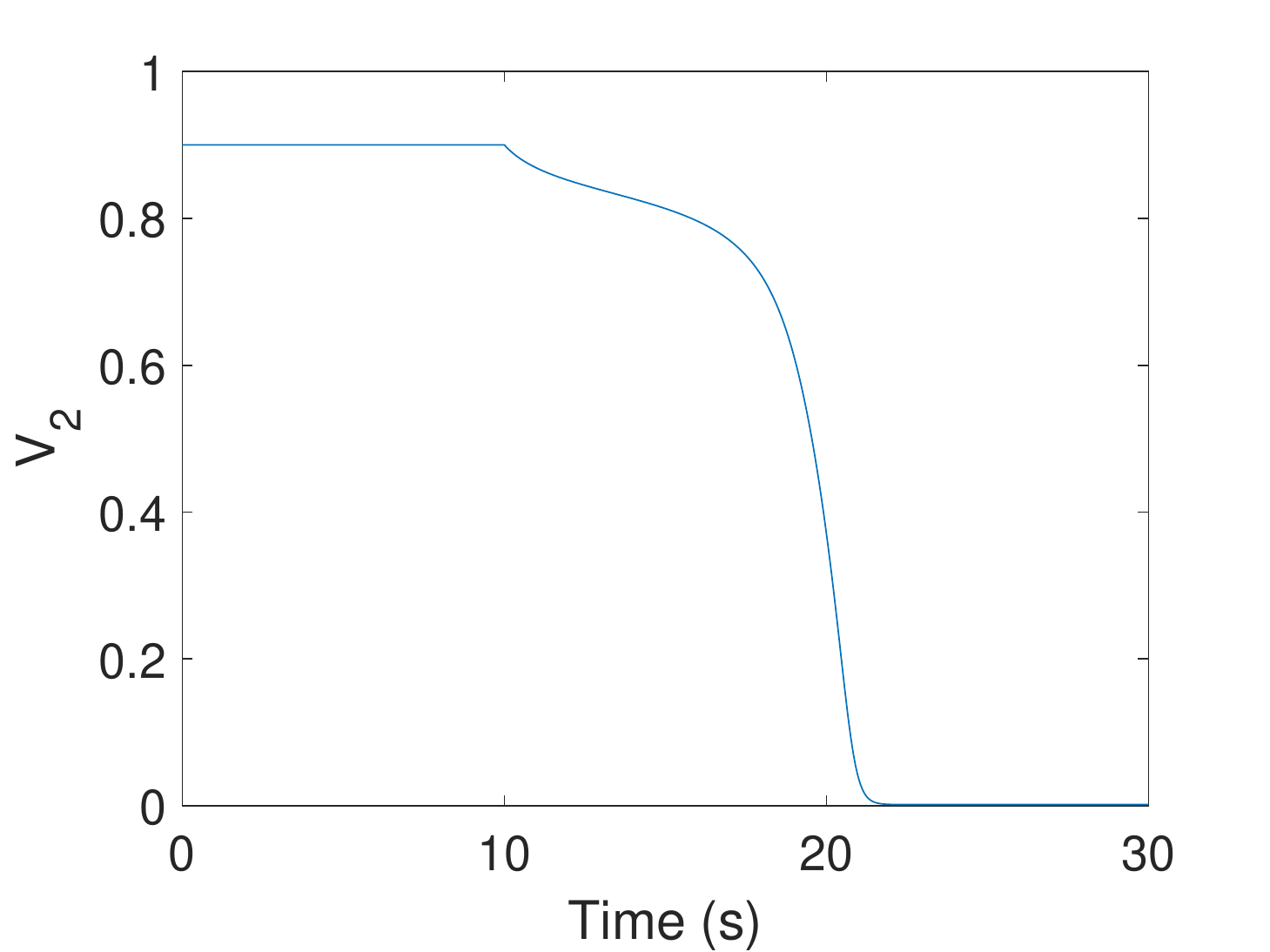}%
\label{TD_dist_time_series}}
\caption{Disturbance response}
\label{fig:TD_dist}
\end{figure}

In Fig.~\ref{fig:TD_dist}, we present simulation results of the benchmark system's disturbance response, again using the parameters in Table \ref{tab:TD_parameters}. A disturbance to both the source voltage $V_1$ and line resistance $R$ causes the system's P-V curve to shrink inwards. This results in the loss of the high voltage stable equilibrium point. Consequently, the voltage decays to the new low voltage stable equilibrium point. Not only do these results demonstrate what we might call `complete' voltage collapse, but the model uses physical state variables and has a physical interpretation, which is useful.

\subsection{DL: Simulation results}
\begin{figure}[!h]
\centering
\subfloat[Maximum root driven trajectories]{\includegraphics[width=.8\linewidth]{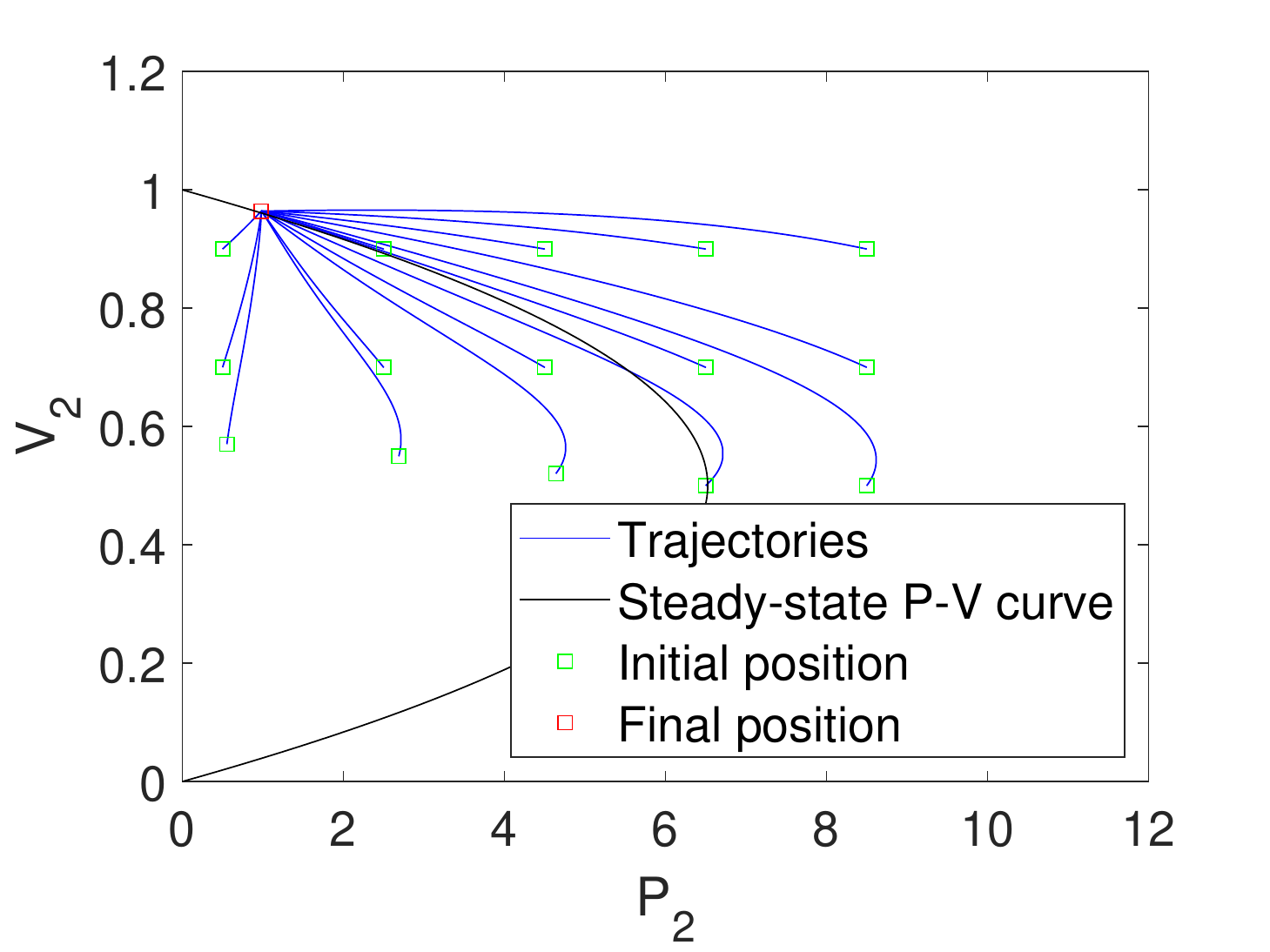}%
\label{DL_high_PV}}
\hfill
\subfloat[Minimum root driven trajectories]{\includegraphics[width=.8\linewidth]{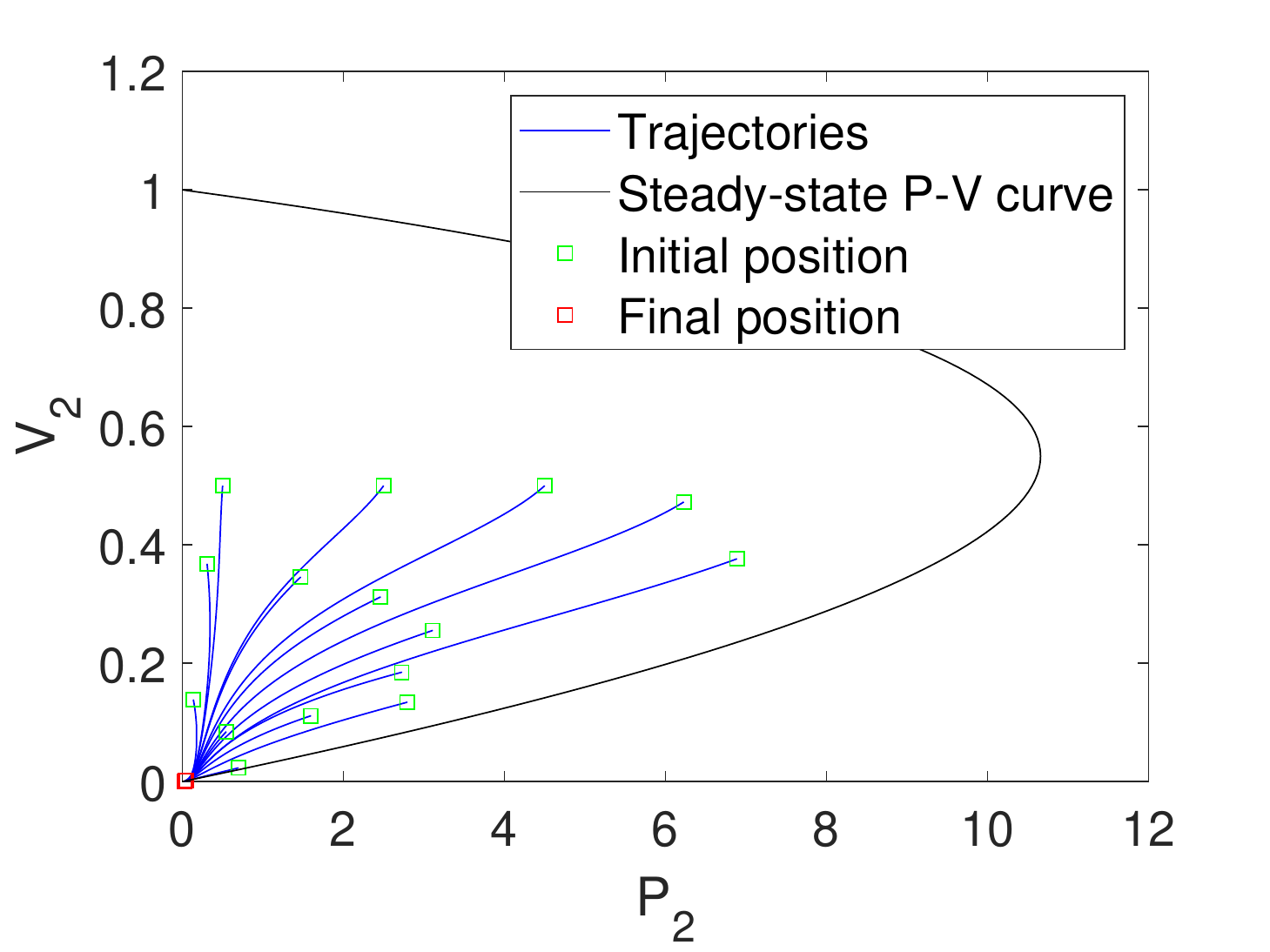}%
\label{DL_low_PV}}
\caption{DL model $P$--$V$ trajectories under two different modelling choices regarding \eqref{eqn:eq1}, using the same set of initial load states $x$ and $y$. The system has two equilibrium points, at $V_2=0.9612$ and $V_2=0.0002889$.}
\label{fig:DL_trajs}
\end{figure}

We present our simulation results that investigate the system response to a range of initial load states, $x$ and $y$. In Fig~\ref{fig:DL_trajs}, we present our simulation results obtained using the parameters displayed in Table~\ref{tab:DL_parameters}. 

\begin{table}[!h]
    \centering
    \caption{DL model parameters}
    \label{tab:DL_parameters}
    \begin{tabular}{cc}
    \hline
        Parameter & Value \\ \hline
        $V_1$ & 1 \\ 
        R    & 0.02 \\ 
        X    &  0.02 \\ 
        $T_p$ & 1 \\ 
        $T_q$ & 1 \\ 
        a     & 0.5625 \\ 
        b & 3 \\ 
        $Q_0$ & 1 \\ 
        $P_0$ & 1  \\ \hline
 \end{tabular}
\end{table}

Specifying initial states $x$ and $y$ does not specify a unique system voltage and power, due to \eqref{eqn:eq1} having multiple voltage solutions. Therefore, in addition to specifying initial states $x$ and $y$, we also have to select which solution of \eqref{eqn:eq1} to use. If we select the high-voltage solution at each timestep, then we get trajectories that converge to the low voltage equilibrium at $V_2=0.0002889$ (see Fig.~\ref{DL_high_PV}), and if we select the low-voltage solution at each timestep, we get trajectories that converge to the high voltage equilibrium at $V_2=0.9612$ (see Fig.~\ref{DL_low_PV}). Therefore, this model is limited if we want to investigate its dynamic response from all possible initial conditions. \\

\begin{figure}[!h]
\centering
\subfloat[Case 1: Two post-disturbance equilibrium points.]{\includegraphics[width=.8\linewidth]{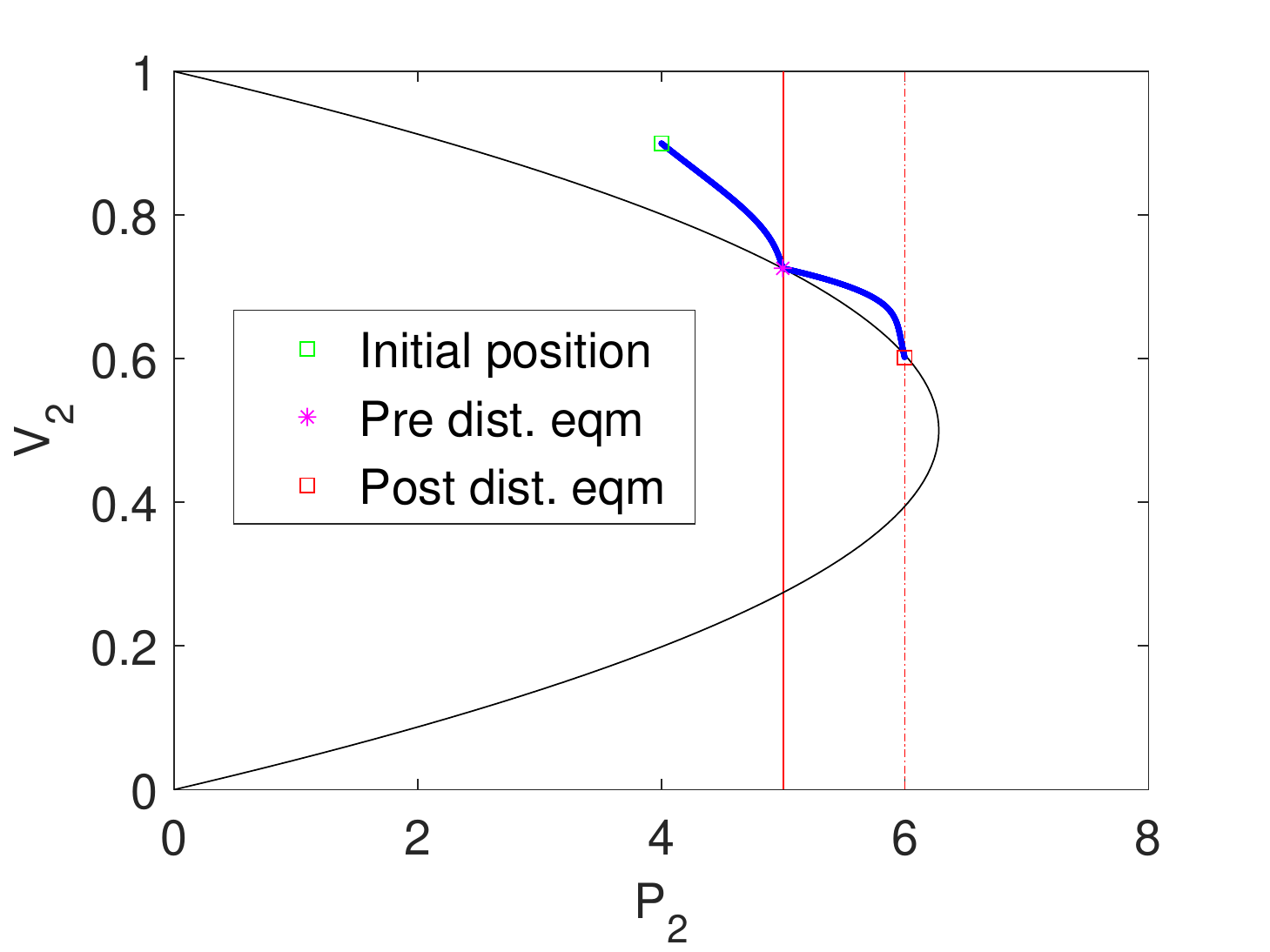}%
\label{DLR_dist_2a}}
\hfill
\subfloat[Case 2: One post-disturbance equilibrium point at $V_2=0$.]{\includegraphics[width=.8\linewidth]{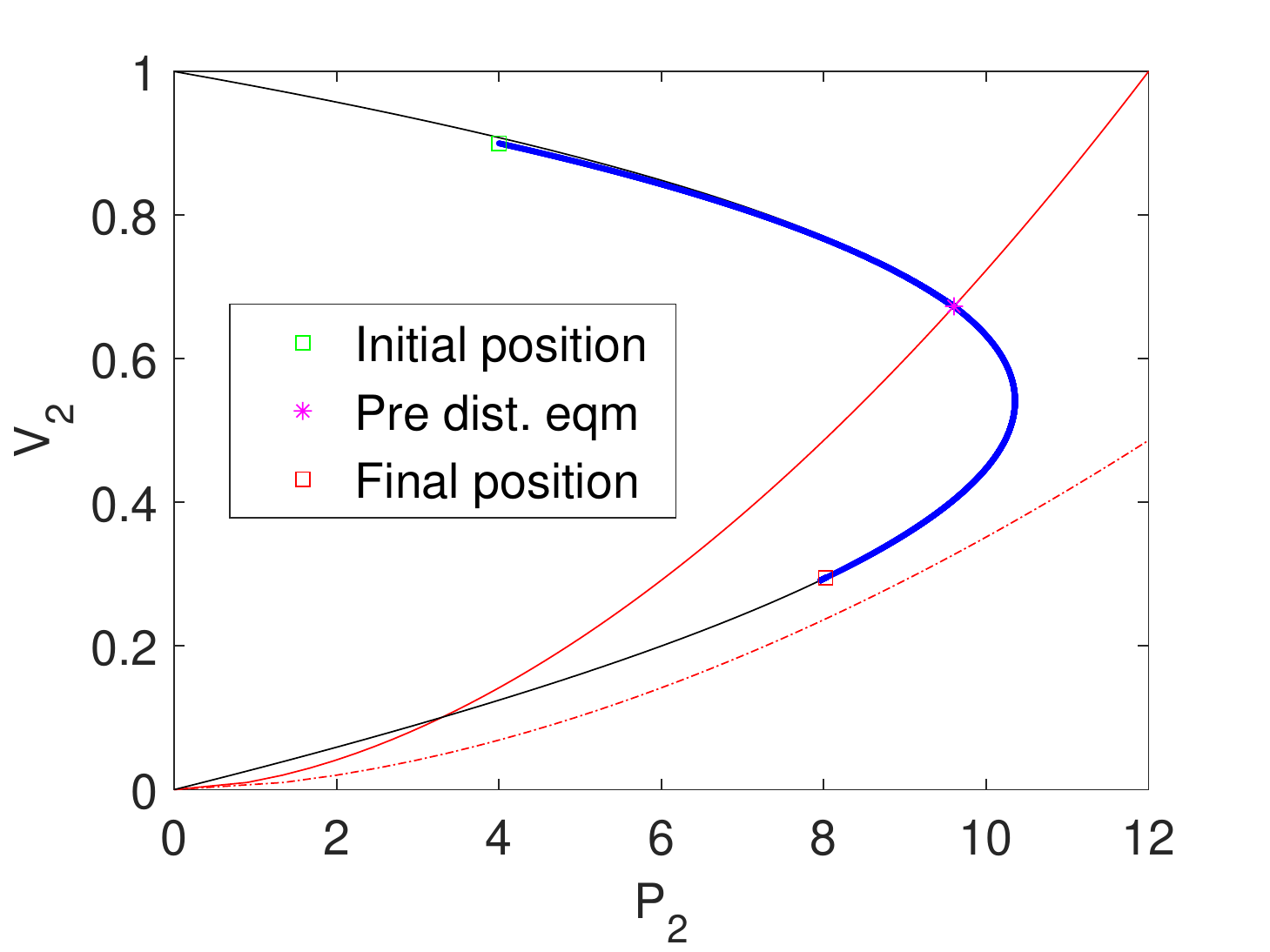}%
\label{DLR_dist_1a}}
\hfill
\subfloat[Case 3: No post-disturbance equilibrium points.]{\includegraphics[width=.8\linewidth]{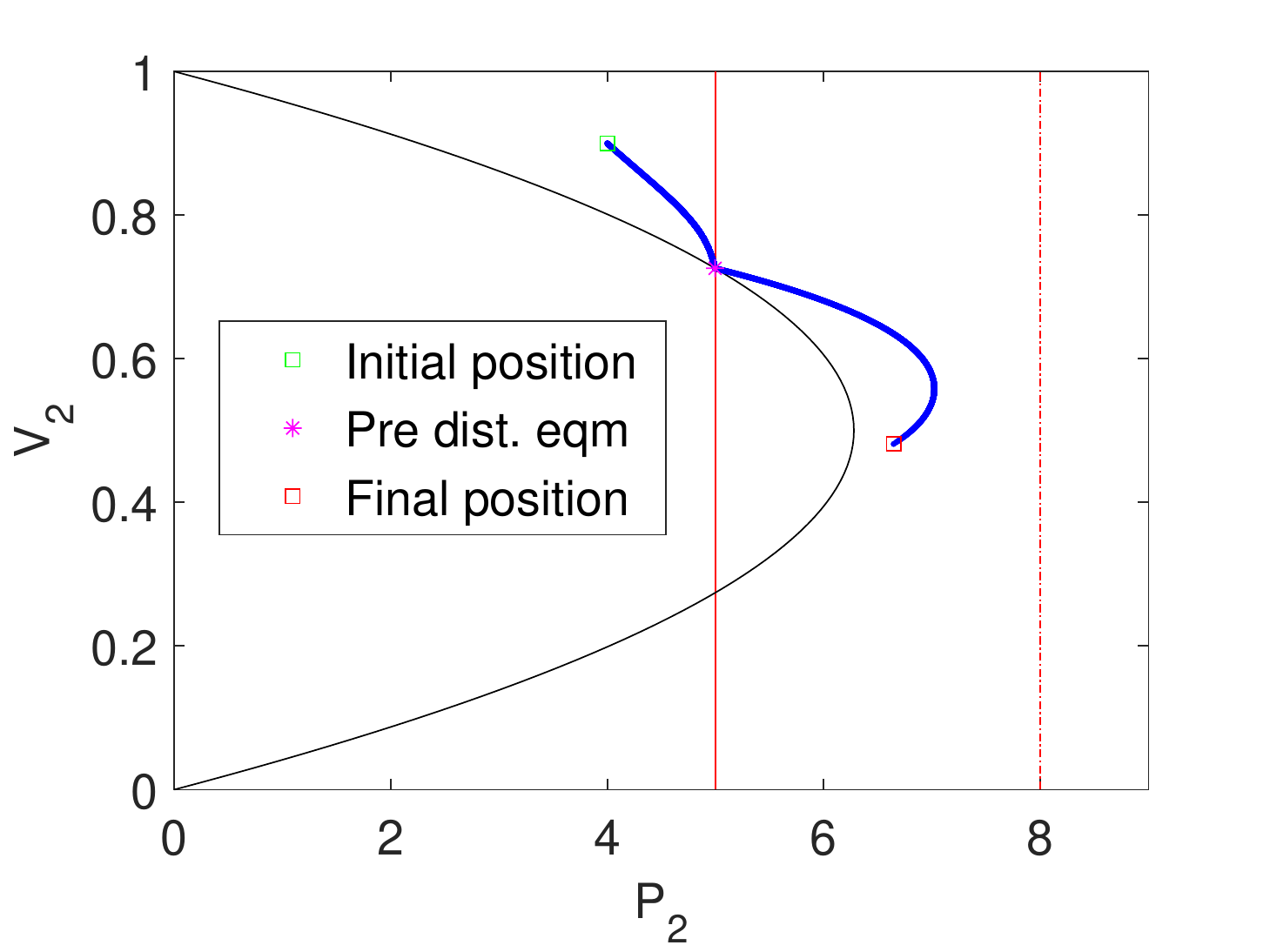}%
\label{DLR_dist_0a}}
\caption{DL system trajectories (blue) in response to load disturbance, with P-V curve (black), pre-disturbance $P_s$ (solid red), and post-disturbance $P_s$ (dashed red).}
\label{fig:DLR_dist_trajs}
\end{figure}

We simulate the system response to a load level increase by applying a step increase to $P_0$ and $Q_0$. We let the maximum voltage solution drive the simulation for three distinct cases. In Case 1, the post-disturbance system still has two equilibrium voltages. In Case 2, there is only one post-disturbance equilibrium voltage, at $V_2=0$. In Case 3, there are no post-disturbance equilibrium points. The parameters for each case are presented in Table~\ref{tab:DLR_dist_params}. In Fig.~\ref{fig:DLR_dist_trajs} we present the simulation results. \\

\begin{table}[!h]
    \centering
    \caption{Model parameters}
    \label{tab:DLR_dist_params}
    \begin{tabular}{cccc}
    \hline
        \multirow{2}{4em}{Parameter} & Case 1 & Case 2 & Case 3 \\  
         & 2 post-dist. eqm & 1 post-dist. eqm & 0 post-dist. eqm \\ \hline
        $V_1$ & 1 & 1 & 1 \\ 
        R    & 0.02 & 0.02 & 0.02 \\ 
        X    & 0.02 & 0.02 & 0.02 \\ 
        $T_p$ & 1 & 1 & 1 \\ 
        $T_q$ & 1 & 1 & 1 \\ 
        a     & 0 & 0.5625 & 0 \\ 
        b     & 0 & 3 & 0\\ 
        $Q_0$ & 5 & 0 & 5\\ 
        $P_0$ & 5 & 12 & 5\\ 
        $Q_0$ step & 1 & 0 & 3  \\ 
        $P_0$ step & 1 & 6 & 3\\ \hline
    \end{tabular}
\end{table}

In Fig.~\ref{DLR_dist_2a}, we observe that if the post-disturbance system still has a high voltage equilibrium, then the load dynamics move the system to that new operating point. \\ 

In Fig.~\ref{DLR_dist_1a}, we observe that even when we can analytically calculate a single equilibrium point at $V_2=0$, we cannot reach it. The simulation stops prematurely because the system reaches a point where there are no longer real, positive voltage solutions to the algebraic constraint. \\

In Fig.~\ref{DLR_dist_0a}, we observe after the disturbance, the load dynamics pull the system voltage down, but the simulation ends prematurely because the system reaches a point where there are no longer real, positive voltage solutions to the algebraic constraint. Again, this result is unsatisfactory because it doesn't indicate whether the voltage goes to $0$, or what behaviour occurs beyond this point. Furthermore, the use of abstract states $x$ and $y$ rather than physical states like $V$, $P$, and $Q$ makes providing a physical interpretation of the results more difficult. \\

\begin{figure}[!h]
\centering
    \includegraphics[width=.75\linewidth]{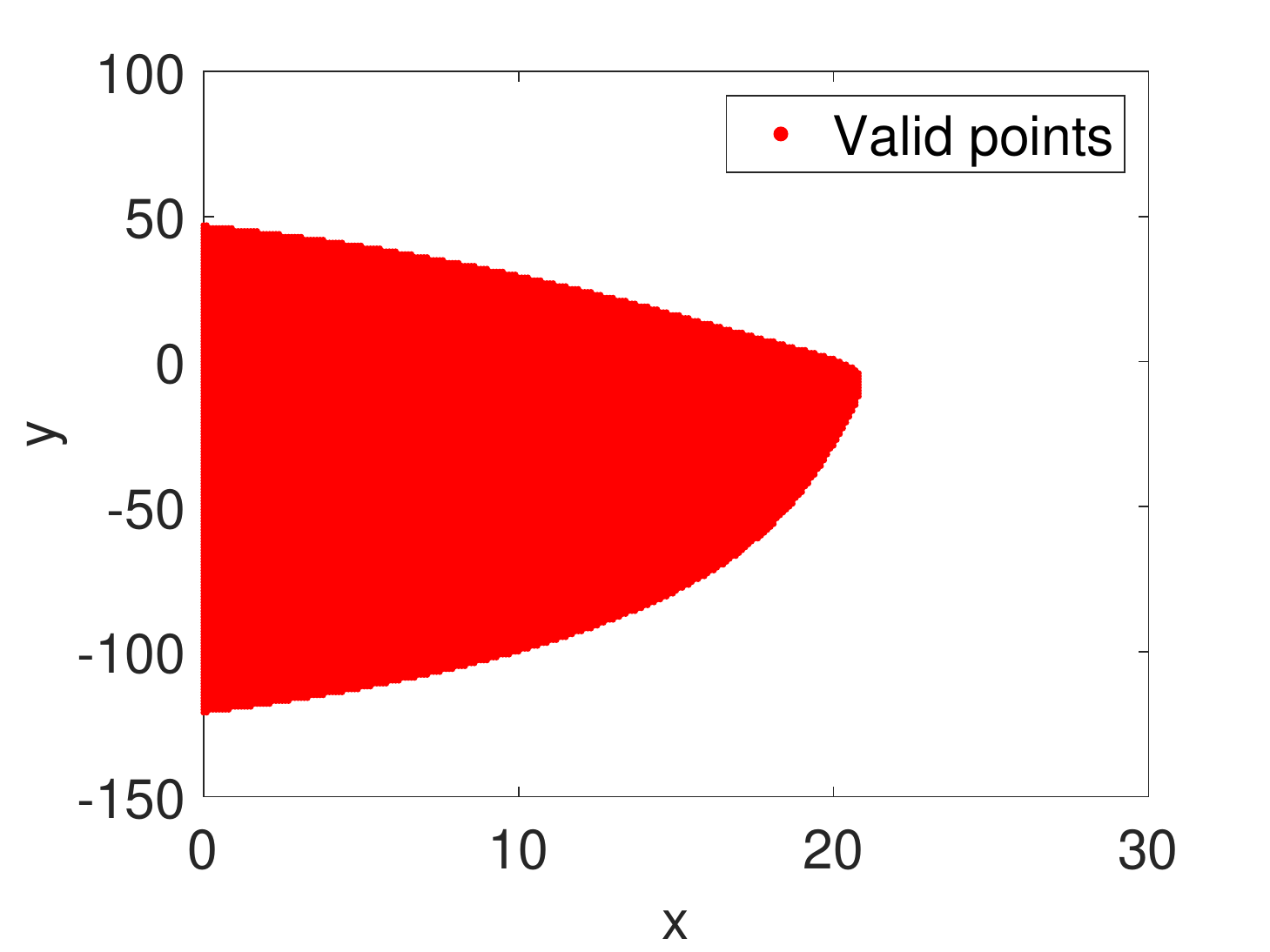}
    \caption{Valid points ($x$--$y$ pairs) in load state space that give two positive real voltage solutions to \eqref{eqn:eq1}, assuming $V_2\geq0$ and $P_2\geq0$}
    \label{fig:xyregions}
\end{figure}

Lastly, we examine the load state space for the system with parameters given in Table \ref{tab:DL_parameters} and observe that if there exist two positive real solutions to \eqref{eqn:eq1_load_eqm}, and we use initial states $x$ and $y$ that lie within the valid region shown in Fig.~\ref{fig:xyregions}, the system will always converge to either the high voltage equilibrium (if driven by the maximum voltage solution) or the low voltage equilibrium (if driven by the minimum voltage solution). In this sense, each equilibrium point is stable. However, this feature is likely to be a product of the specific choices for \eqref{eqn:Ps}--\eqref{eqn:Qt}. Alternative load characteristics should be simulated to determine the conditions under which unstable equilibrium voltages could exist in this system.    

The fundamental limitation of the DL model is its inability to model behaviour across the entire power-voltage domain. By comparing the DL model to the benchmark model, we can gain important insight into the study of load dynamics and their impact on voltage stability in distribution networks. First, we observed that the concepts of stable equilibria and regions of attraction from control theory applicable to power systems. The DL model has a `stable' operating region, which is the region where we get two real positive voltage solutions to \eqref{eqn:eq1}. However, we cannot simulate trajectories that enter or exit this region, and therefore we cannot improve our understanding of what physically happens in the `unstable' regions. This contrasts with the benchmark model, where the load dynamics actually move the system from an acceptable operating point to another, unacceptable operating point. 

\section{Conclusion}\label{sec:conclusion}

In this paper we have demonstrated, via simulation, that the DL model has limitations with respect to the benchmark model. The non-steady state simulations and disturbance response simulations indicate that the load dynamics themselves do not cause voltage instability or collapse. Furthermore, the load model does not improve our understanding of voltage stability in distribution networks, since we cannot observe the system's behaviour across the entire power-voltage domain under a single set of modelling choices. Overall, our simulation results have highlighted that the load model choice has a quantitative influence on voltage stability studies, but importantly it also defines the possible voltage trajectories within the system. Future work developing new load models that provide a more physical representation of voltage instability and collapse in distribution networks is possible.

\bibliographystyle{IEEEtran}

\begin{thebibliography}{10}
\providecommand{\url}[1]{#1}
\csname url@samestyle\endcsname
\providecommand{\newblock}{\relax}
\providecommand{\bibinfo}[2]{#2}
\providecommand{\BIBentrySTDinterwordspacing}{\spaceskip=0pt\relax}
\providecommand{\BIBentryALTinterwordstretchfactor}{4}
\providecommand{\BIBentryALTinterwordspacing}{\spaceskip=\fontdimen2\font plus
\BIBentryALTinterwordstretchfactor\fontdimen3\font minus
  \fontdimen4\font\relax}
\providecommand{\BIBforeignlanguage}[2]{{%
\expandafter\ifx\csname l@#1\endcsname\relax
\typeout{** WARNING: IEEEtran.bst: No hyphenation pattern has been}%
\typeout{** loaded for the language `#1'. Using the pattern for}%
\typeout{** the default language instead.}%
\else
\language=\csname l@#1\endcsname
\fi
#2}}
\providecommand{\BIBdecl}{\relax}
\BIBdecl

\bibitem{IEEE_CIGRE_Definitions}
P.~{Kundur}, J.~{Paserba}, V.~{Ajjarapu}, G.~{Andersson}, A.~{Bose},
  C.~{Canizares}, N.~{Hatziargyriou}, D.~{Hill}, A.~{Stankovic}, C.~{Taylor},
  T.~{Van Cutsem}, and V.~{Vittal}, ``Definition and classification of power
  system stability ieee/cigre joint task force on stability terms and
  definitions,'' \emph{IEEE Transactions on Power Systems}, vol.~19, no.~3, pp.
  1387--1401, 2004.

\bibitem{Kundar_textbook}
P.~{Kundar}, \emph{Power System Stability and Control}.\hskip 1em plus 0.5em
  minus 0.4em\relax McGraw-hill, 1994, vol.~9.

\bibitem{1994_Overbye}
T.~Overbye, ``Effects of load modelling on analysis of power system voltage
  stability,'' \emph{International Journal of Electrical Power \& Energy
  Systems}, vol.~16, no.~5, pp. 329 -- 338, 1994.

\bibitem{Borghetti1997}
A.~Borghetti, R.~Caldon, A.~Mari, and C.~Nucci, ``On dynamic load models for
  voltage stability studies,'' \emph{IEEE Transactions on Power Systems},
  vol.~12, no.~1, pp. 293--303, 1997.

\bibitem{Xu}
W.~{Xu} and Y.~{Mansour}, ``Voltage stability analysis using generic dynamic
  load models,'' \emph{IEEE Transactions on Power Systems}, vol.~9, no.~1, pp.
  479--493, 1994.

\bibitem{Hill_LoadModel}
D.~J. {Hill}, ``Nonlinear dynamic load models with recovery for voltage
  stability studies,'' \emph{IEEE Transactions on Power Systems}, vol.~8,
  no.~1, pp. 166--176, 1993.

\bibitem{MKPal_stability_load}
M.~K. {Pal}, ``Voltage stability conditions considering load characteristics,''
  \emph{IEEE Transactions on Power Systems}, vol.~7, no.~1, pp. 243--249, 1992.

\bibitem{Hiskens_tools_nonlinearities}
I.~Hiskens, ``Analysis tools for power systems—contending with
  nonlinearities,'' \emph{Proceedings of the IEEE}, vol.~83, pp. 1573 -- 1587,
  12 1995.

\bibitem{PBC_energies}
A.~von Meier, E.~L. Ratnam, K.~Brady, K.~Moffat, and J.~Swartz, ``Phasor-based
  control for scalable integration of variable energy resources,''
  \emph{Energies}, vol.~13, no.~1, p. 190, Jan 2020.

\bibitem{Hill_dynamic_analysis_voltage_collapse}
D.~J. {Hill} and I.~A. {Hiskens}, ``Dynamic analysis of voltage collapse in
  power systems,'' in \emph{[1992] Proceedings of the 31st IEEE Conference on
  Decision and Control}, 1992, pp. 2904--2909 vol.3.

\bibitem{khalil2002nonlinear}
H.~Khalil, \emph{Nonlinear Systems}, ser. Pearson Education.\hskip 1em plus
  0.5em minus 0.4em\relax Prentice Hall, 2002.

\end{thebibliography}
%



\end{document}